\documentclass[conference]{IEEEtran}

\usepackage[utf8]{inputenc}
\usepackage[T1]{fontenc}
\usepackage{microtype}
\usepackage{amssymb}
\usepackage{amsmath, caption}
\usepackage{mathtools}
\usepackage{booktabs}
\usepackage[inline]{enumitem}
\usepackage[linesnumbered,algoruled]{algorithm2e}
\usepackage{array}
\usepackage{cite}
\usepackage{subcaption} 
\usepackage{tikz}
\usepackage{pgfplots}
\usepackage[hidelinks]{hyperref}

\usetikzlibrary{arrows, chains, matrix, shapes}
\tikzset{>=stealth}

\usepgfplotslibrary{groupplots}

\newtheorem{observation}{Observation}{\bfseries}{\normalfont}
\newtheorem{lemma}{Lemma}{\bfseries}{\normalfont}
{\bfseries}{\normalfont}
\newtheorem{theorem}{Theorem}{\bfseries}{\normalfont}
\newtheorem{remark}{Remark}{\bfseries}{\normalfont}

\newcolumntype{L}{>{\centering\arraybackslash}m{2cm}}
\DeclareMathOperator{\X}{\mathsf{X}}
\DeclareMathOperator{\U}{\mathsf{U}}
\DeclareMathOperator{\F}{\mathsf{F}}
\DeclareMathOperator{\G}{\mathsf{G}}
\newcommand{\True}{1}

\begin{document}

\title{Learning Linear Temporal Properties}

\author{
	\IEEEauthorblockN{Daniel Neider}
	\IEEEauthorblockA{Max Planck Institute for Software Systems \\ 67663 Kaiserslautern, Germany \\ Email: neider@mpi-sws.org}
	\and
	\IEEEauthorblockN{Ivan Gavran}
	\IEEEauthorblockA{Max Planck Institute for Software Systems \\ 67663 Kaiserslautern, Germany \\ Email: gavran@mpi-sws.org}
}

\maketitle

\begin{abstract}
We present two novel algorithms for learning formulas in Linear Temporal Logic (LTL) from examples.
The first learning algorithm reduces the learning task to a series of satisfiability problems in propositional Boolean logic and  produces a smallest LTL formula (in terms of the number of subformulas) that is consistent with the given data.
Our second learning algorithm, on the other hand, combines the SAT-based learning algorithm with classical algorithms for learning decision trees. The result is a learning algorithm that scales to real-world scenarios with hundreds of examples, but can no longer guarantee to produce minimal consistent LTL formulas.
We compare both learning algorithms and demonstrate their performance on a wide range of synthetic benchmarks. Additionally, we illustrate their usefulness on the task of understanding executions of a leader election protocol.
\end{abstract}

\section{Introduction}

Making sense of the observed behavior of complex systems is an important problem in practice.
It arises, for instance, in debugging (especially in the context of distributed systems), reverse engineering (e.g., of malware and viruses), specification mining for formal verification, and modernization of legacy systems, to name but a few examples.
However, understanding a system based on examples of its execution is clearly a challenging task that can quickly become overwhelming without proper tool support.

In this paper, we address this problem and develop learning-based techniques to help engineers understand the dynamic (i.e., temporal) behavior of complex systems.
More precisely, we solve the problem of learning formulas in Linear Temporal Logic (LTL)~\cite{DBLP:conf/focs/Pnueli77}, which are meant to distinguish between desirable and undesirable executions of a system (e.g., to explain the root-cause of a bug).
The particular choice of LTL in this work is motivated by two observations: first, logical formulas often provide concise descriptions of the observed behavior and are relatively easy for humans to comprehend; second, LTL---together with Computational Tree Logic (CTL)~\cite{DBLP:conf/lop/ClarkeE81}---is widely considered to be the de~facto standard for specifying temporal properties and, hence, many engineers are familiar with its use.

The precise problem we are aiming at is the following: given a sample $\mathcal S$ consisting of two finite sets of positive and negative examples, learn an LTL formula $\varphi$ that is consistent with $\mathcal S$ in the sense that all positive examples satisfy $\varphi$, whereas all negative examples violate $\varphi$.\footnote{Note that, in contrast to classical computational learning theory~\cite{DBLP:journals/cacm/Valiant84} and modern statistical machine learning~\cite{foundations-of-data-science-2,DBLP:books/daglib/0087929}, we seek to learn a formula that does not make mistakes on the examples. In fact, separation problems of this sort are of great interest in automata and formal language theory. Prominent examples in this area are the minimization of incompletely-specified state machines~\cite{DBLP:journals/tc/Pfleeger73,DBLP:conf/atva/Neider12} and Regular Model Checking~\cite{DBLP:conf/cav/BouajjaniJNT00,DBLP:conf/nfm/NeiderJ13}.}
To be as general and succinct as possible, we here consider examples to be infinite, ultimately periodic words (e.g., traces of a non-terminating system) and assume the standard syntax of LTL. However, our techniques can easily be adapted to the case of finite words and extend smoothly to arbitrary future-time temporal operators, such as ``release'', ``weak until'', and so on.
We fix all necessary definitions and notations in Section~\ref{sec:preliminaries}.

The main contribution of this work are \textbf{two novel learning algorithms for LTL formulas from data}, one based on SAT solving, the other on learning decision trees. \par\medskip

\paragraph*{SAT-based learning algorithm}
The idea of our first algorithm, presented in Section~\ref{sec:sat}, is to reduce the problem of learning an LTL formula to a series of satisfiability problems in propositional Boolean logic and to use highly-optimized SAT solvers to search for solutions.
Inspired by ideas from bounded model checking~\cite{DBLP:journals/ac/BiereCCSZ03}, our learning algorithm produces a series of propositional formulas $\Phi_n^\mathcal S$ for increasing values of $n \in \mathbb N \setminus \{ 0 \}$ that depend on the sample $\mathcal S$ and have the following two properties:
\begin{enumerate*}[label={(\arabic*)}]
	\item $\Phi_n^\mathcal S$ is satisfiable if and only if there exists an LTL formula of size $n$ (i.e., with $n$ subformulas) that classifies the examples correctly, and
	\item a model of $\Phi_n^\mathcal S$ contains sufficient information to construct such an LTL formula.
\end{enumerate*}
By increasing the value of $n$ until $\Phi_n^\mathcal S$ becomes satisfiable, we obtain an effective algorithm that learns an LTL formula that is guaranteed to classify the examples correctly (given that the sample is non-contradictory).

By design, our SAT-based learning algorithm has three distinguished features, which we believe are essential in practice.
First, our algorithm learns LTL formulas of minimal size (i.e., with the minimal number of subformulas). 
As we seek to learn formulas to be read by humans, the size of the learned formula is a crucial metric  since larger formulas are generally harder to understand than smaller ones.
Second, once an LTL formula has been learned, our algorithm can be queried for further, distinct formulas that are consistent with the sample. We believe that this feature is important in practice as it allows generating multiple explanations for the observed data.
Third, our algorithm does not rely on an a~priori given set of templates, which is in stark contrast to existing work on learning temporal properties (e.g., Bombara et~al.~\cite{DBLP:conf/hybrid/BombaraVPYB16}). To the best of our knowledge, our SAT-based algorithm is in fact the first learning algorithm that is not restricted to a fixed class of templates. However, restrictions to the shape of LTL formulas (e.g., to the popular GR(1)-fragment of LTL~\cite{DBLP:journals/jcss/BloemJPPS12}) can easily be encoded if desired. \par\medskip

\paragraph*{Learning algorithm based on decision trees}
Our second learning algorithm, which we present in Section~\ref{sec:dtMethod}, trades in the guarantee of finding minimal solutions in order to attain better scalability.
The key idea is to perform the learning in two phases. In the first phase, we run the SAT-based learning algorithm described above on various subsets of the examples.
This results in a (small) number of LTL formulas, named ``{LTL primitives}'', that classify at least these subsets correctly.
In the second phase we use a standard learning algorithm for decision trees~\cite{DBLP:books/mk/Quinlan93} to learn a Boolean combination of these LTL primitives that classifies the whole set of examples correctly, though it might not be minimal.
Note, however, that we need to carefully choose the subsets of examples such that the resulting LTL primitives
\begin{enumerate*}[label={(\alph*)}]
	\item separate all pairs of positive and negative examples and
	\item are general enough to permit ``small'' decision trees.
\end{enumerate*}
We have experimented with numerous strategies to select subsets, but in this paper we present only the two that performed best.
A well known advantage of decision trees is that they are simple to comprehend due to their rule-based structure.
\par\medskip

In Section~\ref{sec:evaluation}, we evaluate the performance of both learning algorithms on a wide range of synthetic benchmarks that reflect typical patterns of LTL formulas used in practice. Additionally, we illustrate their usefulness for understanding causes of inconsistencies in the leader election used by Zookeeper's atomic broadcast protocol~\cite{DBLP:conf/dsn/JunqueiraRS11}.


\subsection*{Related Work}

Learning of temporal properties from examples has recently attracted increasing interest, especially in the area of \emph{Signal Temporal Logic (STL)}~\cite{DBLP:conf/formats/MalerN04} and \emph{parametric STL}~\cite{DBLP:conf/rv/AsarinDMN11}.
Examples include the work by Asarin et al.~\cite{DBLP:conf/rv/AsarinDMN11}, Kong et al.~\cite{DBLP:conf/hybrid/KongJAGB14,DBLP:journals/tac/KongJB17}, Vaidyanathan et al.~\cite{DBLP:conf/cdc/VaidyanathanIBD17}, and Bartocci, Bortolussi, and Sanguinetti~\cite{DBLP:journals/corr/BartocciBS13}.
In contrast to our SAT-based learning algorithm, however, all of these techniques either rely on user-given templates or can only learn formulas from very restricted syntactic fragments.
Various techniques for mining LTL specifications~\cite{DBLP:conf/memocode/LiDS11,DBLP:conf/kbse/LemieuxPB15} and CTL specifications~\cite{DBLP:journals/ase/WasylkowskiZ11} exist as well, but these also rely on templates or restrict the class of formulas severely.
To the best of our knowledge, our SAT-based algorithm is in fact the first that is capable of learning unrestricted LTL formulas without relying on user-given templates.
Nonetheless, expert knowledge in form of constraints on the syntax can easily be encoded if desired.

Our SAT-based learning algorithm is inspired by bounded model checking~\cite{DBLP:journals/ac/BiereCCSZ03} and earlier work of the first author on learning (minimal) automata over finite words~\cite{DBLP:conf/atva/Neider12,DBLP:conf/nfm/NeiderJ13}.
However, since regular languages are strictly more expressive than LTL (the former being equivalent to monadic second-order logic~\cite{Buechi}, while the latter being equivalent to fist-order logic~\cite{Kamp1968-KAMTLA}), automata learning techniques---including active learning algorithms~\cite{DBLP:conf/tacas/FarzanCCTW08,DBLP:journals/tcs/AngluinF16} that operate in Angluin's active learning framework~\cite{DBLP:journals/iandc/Angluin87}---are not immediately applicable.
However, lifting the methods developed in this work to an active learning setup, without a detour via automata, is part of our plans for future work.

Using decision trees to learn Signal Temporal Logic (STL) formulas has been explored by Bombara et al.~\cite{DBLP:conf/hybrid/BombaraVPYB16}, whose main contribution is an adaptation of the classical impurity measure to account for STL formulas.
However, this work still requires user-defined STL primitives to be provided, which serve as the features for the decision tree learning algorithm.
By contrast, our technique uses the SAT-based learning algorithm to infer LTL primitives fully automatically.

Learning of logical formulas has also been studied in the context of \emph{probably approximately correct learning (PAC)}~\cite{DBLP:journals/cacm/Valiant84}.
Grohe and Ritzert~\cite{DBLP:conf/lics/GroheR17}, for instance, considered learning of first-order definable concepts over structures of small degree. Subsequently, Grohe, Löding, and Ritzert~\cite{DBLP:conf/alt/GroheLR17} studied the learning of hypotheses definable using monadic second order logic on strings.
Due to the fundamental differences between PAC learning and the learning model considered here (one being approximate and the other being exact), their techniques cannot easily be applied.

\section{Preliminaries}
\label{sec:preliminaries}
In this section, we set up definitions and notations used throughout the paper.

\subsubsection*{Finite and Infinite Words}
An \emph{alphabet} $\Sigma$ is a nonempty, finite set. The elements of this set are called \emph{symbols}.

A \emph{finite word} over an alphabet $\Sigma$ is a sequence $u = a_0 \ldots a_n$ of symbols $a_i \in \Sigma$, $i \in \{ 0, \ldots, n \}$. The empty sequence is called \emph{empty word} and written as $\varepsilon$.
The length of a finite word $u$ is denoted by $|u|$, where $|\varepsilon| = 0$.
Moreover, $\Sigma^\ast$ denotes the set of all finite words over the alphabet $\Sigma$, while $\Sigma^+ = \Sigma^\ast \setminus \{ \varepsilon \}$ is the set of all non-empty words. 

An \emph{infinite word} over $\Sigma$ is an infinite sequence $\alpha = a_0 a_1 \ldots $ of symbols $a_i \in \Sigma$, $i \in \mathbb N$. We denote the $i$-th symbol of an infinite word $\alpha$ by $\alpha(i)$ and the infinite suffix starting at position $j$ by $\alpha[j, \infty)$.
Given $u \in \Sigma^+$, the infinite word $u^\omega = u u \ldots \in \Sigma^\omega$ is the infinite repetition of $u$.
An infinite word $\alpha$ is called \emph{ultimately periodic} if it is of the form $\alpha = uv^\omega$ for a $u \in \Sigma^\ast$ and $v \in \Sigma^+$.
Finally, $\Sigma^\omega$ denotes the set of all infinite words over the alphabet $\Sigma$. 

\subsubsection*{Propositional Boolean Logic}
Let $\mathit{Var}$ be a set of propositional variables, which take Boolean values from $\mathbb B = \{ 0, 1 \}$ ($0$ representing $\mathit{false}$ and $1$ representing $\mathit{true}$). Formulas in \emph{propositional (Boolean) logic}---which we denote by capital Greek letters---are inductively constructed as follows: 
\begin{itemize}
	\item each $x \in \mathit{Var}$ is a propositional formula; and
	\item if $\Psi$ and $\Phi$ are propositional formulas, so are $\lnot \Psi$ and $\Psi \lor \Phi$.
\end{itemize}
Moreover, we add syntactic sugar and allow the formulas $\mathit{true}$, $\mathit{false}$, $\Psi \land \Phi$, $\Psi \Rightarrow \Phi$, and $\Psi \Leftrightarrow \Phi$, which are defined as usual.

A \emph{propositional valuation} is a mapping $v \colon \mathit{Var} \to \mathbb B$, which maps propositional variables to Boolean values. 
The semantics of propositional logic is given by a satisfaction relation $\models$ that is inductively defined as follows: $v \models x$ if and only if $v(x) = 1$, $v \models \lnot \Psi$ if and only if $v \not\models \Psi$, and $v \models \Psi \lor \Phi$ if and only if $v \models \Psi$ or $v \models \Phi$.
In the case that $v \models \Phi$, we say that \emph{$v$ satisfies $\Phi$} and call it a \emph{model} of $\Phi$.
A propositional formula $\Phi$ is \emph{satisfiable} if there exists a model $v$ of $\Phi$.
The \emph{size} of a formula is the number of its subformulas (as defined in the usual way).

The satisfiability problem of propositional logic is the problem to decide whether a given formula is satisfiable. Although this problem is well-known to be NP-complete~\cite{DBLP:series/faia/2009-185}, modern SAT solvers implement optimized decision procedures that can check satisfiability of formulas with millions of variables~\cite{DBLP:conf/aaai/BalyoHJ17}. Moreover, SAT solvers also return a model if the input-formula is satisfiable.

\subsubsection*{Linear Temporal Logic}
\emph{Linear Temporal Logic (LTL)}~\cite{DBLP:conf/focs/Pnueli77} is an extension of propositional Boolean logic with modalities that allow expressing temporal properties. Starting with a finite, nonempty set $\mathcal P$ of \emph{atomic propositions}, formulas in LTL---which we denote by small Greek letters---are inductively defined as follows:
\begin{itemize}
	\item each atomic proposition $p \in \mathcal P$ is an LTL formula;
	\item if $\psi$ and $\varphi$ are LTL formulas, so are $\lnot \psi$, $\psi \lor \varphi$, $\X \psi$ (``next''), and $\psi \U \varphi$ (``until'').
\end{itemize}
Again, we add syntactic sugar and allow the formulas $\mathit{true} \coloneqq p \lor \lnot p$ for some $p \in \mathcal P$, $\mathit{false} \coloneqq \lnot \mathit{true}$, as well as $\psi \land \varphi$ and $\psi \rightarrow \varphi$, which are defined as usual.
Moreover, we allow the additional temporal formulas $\F \psi \coloneqq \mathit{true} \U \psi$ (``finally'') and $\G \psi \coloneqq \lnot \F \lnot \psi$ (``globally'').
The \emph{size} of an LTL formula $\varphi$, which we denote by $|\varphi|$, is the number of its subformulas.
Finally, let $\mathcal C = \{ \land, \lor, \neg, \rightarrow, \F, \G, \U, \X \}$ be the set of LTL operators.

LTL formulas are interpreted over infinite words $\alpha \in (2^\mathcal P)^\omega$, though there exist various semantics for LTL over finite words and our techniques smoothly extend to these situations.
For the sake of a simpler presentation, we define the semantics of LTL in a slightly non-standard way by means of a \emph{valuation function $V$}. This functions maps pairs of LTL formulas and infinite words to Boolean values and is inductively defined as follows:
$V(p, \alpha) = 1$ if and only if $p \in \alpha(0)$, $V(\lnot \varphi, \alpha) = 1 - V(\varphi, \alpha)$, $V(\varphi \lor \psi, \alpha) = \max{\{ V(\varphi, \alpha), V(\psi, \alpha) \}}$, $V(\X \varphi, \alpha) = V(\varphi, \alpha[1, \infty))$, and $V(\varphi \U \psi, \alpha) = \max_{i \geq 0}{\{ \min{\{ V(\psi, \alpha[i, \infty)), \min_{0 \leq j < i} {\{ V(\varphi, \alpha[j, \infty)) \}} \}} \}}$.
We call $V(\varphi, \alpha)$ the \emph{valuation of $\varphi$ on $\alpha$} and say that \emph{$\alpha$ satisfies $\varphi$} if $V(\varphi, \alpha) = 1$.

Our SAT-Based learning algorithm relies on a canonical syntactic representation of LTL formulas, which we call \emph{syntax DAGs}. A syntax DAG is essentially a syntax tree (i.e., the unique tree labeled with atomic propositions as well as Boolean and temporal operators that is derived from the inductive definition of an LTL formula) in which common subformulas are shared.
This sharing turns the syntax tree into a directed, acyclic graph (DAG), whose number of nodes coincides with the number of subformulas of the represented LTL formula.
As an example, Figure~\ref{subfig:syntax-dag} (on Page~\pageref{fig:trees}) depicts the (unique) syntax DAG of the formula $(p \U {\G q}) \lor (\F\G q)$, in which the subformula $\G q$  is shared; the corresponding syntax tree is depicted in Figure~\ref{subfig:syntax-tree}.
Note that syntactically distinct formulas have different (i.e., non-isomorphic) syntax DAGs.

\subsubsection*{Samples and Consistency}
Throughout this paper, we assume that the data we learn from is given as two (potentially empty) finite, disjoint sets $P, N \subset (2^{\mathcal P})^\omega$ of ultimately periodic words. The words in $P$ are interpreted as \emph{positive examples}, while the words in $N$ are interpreted as \emph{negative examples}. We call the pair $\mathcal S = (P, N)$ a \emph{sample}. 
Since we want to work with the ultimately periodic words in a sample algorithmically, we assume that they are stored as pairs $(u, v)$ of finite words $u \in (2^\mathcal P)^\ast$ and $v \in (2^\mathcal P)^+$, which can be accessed individually.
To measure the complexity of a sample, we define its \emph{size} to be $|\mathcal S| = \sum_{uv^\omega \in P \cup N} |u| + |v|$.

Given an LTL formula $\varphi$ and a sample $\mathcal S = (P, N)$, both over a set $\mathcal P$ of atomic propositions, we call $\varphi$ \emph{consistent} with $\mathcal S$ if $V(\varphi, uv^\omega) = 1$ for each $uv^\omega \in P$ (i.e., all positive examples satisfy $\varphi$) and $V(\varphi, uv^\omega) = 0$ for each $uv^\omega \in N$ (i.e., all negative examples do not satisfy $\varphi$); in this case, we also say that $\varphi$ \emph{separates} $P$ and $N$.
We call $\varphi$ \emph{minimally consistent with $\mathcal S$} if $\varphi$ is consistent with $\mathcal S$ and no consistent LTL formula of smaller size exists.

\section{A SAT-based Learning Algorithm}
\label{sec:sat}
The fundamental task we solve in this section is:
\begin{quote}
	\itshape
	``given a sample $\mathcal S$, compute an LTL formula of minimal size that is consistent with $\mathcal S$''. 
\end{quote}
We call this task \emph{passive learning of LTL formulas}---as opposed to active learning~\cite{DBLP:journals/iandc/Angluin87} where the learning algorithm is permitted to actively query for additional data.
Note that this problem can have more than one solution as there can be multiple, non-equivalent LTL formulas that are minimally consistent with a given sample.

Before we explain our learning algorithm in detail, let us briefly comment on the minimality requirement in the definition above.
On the one hand, we observe that the problem becomes simple if no restriction on the size is imposed:
for $\alpha \in P$ and $\beta \in N$, construct a formula $\varphi_{\alpha, \beta}$ with $V(\varphi_{\alpha, \beta}, \alpha) = 1$ and $V(\varphi_{\alpha, \beta}, \beta) = 0$ that describes the first symbol where $\alpha$ and $\beta$ differ using a sequence of $\X$-operators and an appropriate propositional formula;
then, $\bigvee_{\alpha \in P} \bigwedge_{\beta \in N} \varphi_{\alpha, \beta}$ is consistent with $\mathcal S$ since we assume $P$ and $N$ to be disjoint.
However, simply characterizing all differences between positive and negative examples is clearly overfitting the sample and, hence, arguably of little help in practice.
On the other hand, we believe that small formulas are easier for humans to comprehend than large ones, which justifies spending effort on learning a smallest formula.
However, we do not impose any preference amongst minimal consistent formulas (which is an interesting topic for future work).

Let us now turn to describing our learning algorithm.
Its underlying idea is to reduce the construction of a minimally consistent LTL formula to a satisfiability problem in propositional logic and use a highly-optimized SAT solver to search for solutions. More precisely, given a sample $\mathcal S$ and a natural number $n \in \mathbb N \setminus \{ 0 \}$, we construct a propositional formula $\Phi_n^\mathcal S$ of size polynomial in $n$ and $|\mathcal S|$ that has the following two properties:
\begin{enumerate}
	\item $\Phi_n^\mathcal S$ is satisfiable if and only if there exists an LTL formula of size $n$ that is consistent with $\mathcal S$; and
	\item if $v$ is a model of $\Phi_n^\mathcal S$, then $v$ contains sufficient information to construct an LTL formula $\psi_v$ of size $n$ that is consistent with $\mathcal S$.
\end{enumerate}

By increasing the value of $n$ by one and extracting an LTL formula $\psi_v$ from a model $v$ of $\Phi_n^\mathcal S$ as soon as it becomes satisfiable (indeed, any model is sufficient), we obtain an effective algorithm that learns an LTL formula of minimal size that is consistent with $\mathcal S$.
This idea is shown in pseudo code as Algorithm~\ref{alg:sat-learner}.
In fact, the existence of a trivial solution for the passive LTL learning task (as sketched at the beginning of this section) shows that Algorithm~\ref{alg:sat-learner} is guaranteed to terminate, and the size of this solution provides an upper bound on the value of $n$. 

\begin{algorithm}
	\KwIn{a sample $\mathcal S$}
	\BlankLine
	$n \gets 0$\; 
	\Repeat{$\Phi_n^\mathcal S$ is satisfiable, say with model $v$}
	{
		$n \gets n + 1$\;
		Construct and solve $\Phi_n^\mathcal S$\;
	}
	Construct and \Return $\psi_v$\;

	\caption{SAT-based learning algorithm}\label{alg:sat-learner}
\end{algorithm}

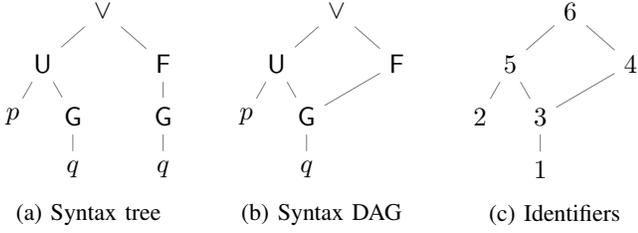
\begin{figure}
	\centering
	\begin{subfigure}{.3\columnwidth}
		\centering
		\begin{tikzpicture}
			\node (1) at (0, 0) {$\lor$};
			\node (2) at (-.8, -.7) {$\U$};
			\node (3) at (.8, -.7) {$\F$};
			\node (4) at (-1.2, -1.4) {$p$};
			\node (5) at (-.4, -1.4) {$\G$};
			\node (6) at (.8, -1.4) {$\G$};
			\node (7) at (-.4, -2.1) {$q$};
			\node (8) at (.8, -2.1) {$q$};
			\draw[gray] (1) -- (2) (1) -- (3) (2) -- (4) (2) -- (5) (3) -- (6) (5) -- (7) (6) -- (8);
		\end{tikzpicture}
		\caption{Syntax tree} \label{subfig:syntax-tree}
	\end{subfigure}
	\hfill
	\begin{subfigure}{.3\columnwidth}
		\centering
		\begin{tikzpicture}
			\node (1) at (0, 0) {$\lor$};
			\node (2) at (-.8, -.7) {$\U$};
			\node (3) at (.8, -.7) {$\F$};
			\node (4) at (-1.2, -1.4) {$p$};
			\node (5) at (-.4, -1.4) {$\G$};
			\node (6) at (-.4, -2.1) {$q$};
			\draw[gray] (1) -- (2) (1) -- (3) (2) -- (4) (2) -- (5) (3) -- (5) (5) -- (6);
		\end{tikzpicture}
		\caption{Syntax DAG} \label{subfig:syntax-dag}
	\end{subfigure}
	\hfill
	\begin{subfigure}{.3\columnwidth}
		\centering
		\begin{tikzpicture}
			\node (1) at (0, 0) {$6$};
			\node (2) at (-.8, -.7) {$5$};
			\node (3) at (.8, -.7) {$4$};
			\node (4) at (-1.2, -1.4) {$2$};
			\node (5) at (-.4, -1.4) {$3$};
			\node (6) at (-.4, -2.1) {$1$};
			\draw[gray] (1) -- (2) (1) -- (3) (2) -- (4) (2) -- (5) (3) -- (5) (5) -- (6);
		\end{tikzpicture}
		\caption{Identifiers} \label{subfig:identifiers}
	\end{subfigure}
	\caption{Syntax tree, syntax DAG, and identifiers of the syntax DAG for the LTL formula $(p \U {\G q}) \lor (\F\G q)$} \label{fig:trees}
\end{figure}

The key idea of the formula $\Phi_n^\mathcal S$ is to encode the syntax DAG of an (unknown) LTL formula $\varphi^\star$ with $n$ subformulas and then constrain the variables of $\Phi_n^\mathcal S$ such that $\varphi^\star$ is consistent with the sample $\mathcal S$.
To simplify our encoding, we assign to each node of this syntax DAG a unique \emph{identifier} $i \in \{ 1, \ldots, n \}$ such that
\begin{enumerate*}[label={(\alph*)}]
	\item the identifier of the root is $n$ and
	\item if the identifier of an inner node is $i$, then the identifiers of its children are less than $i$.
\end{enumerate*}
Note that such a numbering scheme is not unique for a given syntax DAG, but it entails that the root always has identifier $n$ and the node with identifier $1$ is always labeled with an atomic proposition. 
We refer the reader to Figures~\ref{subfig:syntax-dag} and \ref{subfig:identifiers} for an example.

We encode a syntax DAG using three types of propositional variables:
\begin{itemize}
	\item $x_{i, \lambda}$ where $i \in \{ 1, \ldots, n \}$ and $\lambda \in \mathcal P \cup \mathcal C$;
	\item $l_{i, j}$ where $i \in \{ 2, \ldots, n \}$ and $j \in \{ 1, \ldots, i-1 \}$; and
	\item $r_{i, j}$ where $i \in \{ 2, \ldots, n \}$ and $j \in \{ 1, \ldots, i-1 \}$.
\end{itemize}
Intuitively, the variables $x_{i, \lambda}$ encode a labeling of the syntax DAG in the sense that if a variable $x_{i, \lambda}$ is set to $\mathit{true}$, then node $i$ is labeled with $\lambda$ (recall that each node is labeled with either an atomic proposition from $\mathcal P$ or an operator from $\mathcal C$).
The variables $l_{i, j}$ and $r_{i, j}$, on the other hand, encode the structure of the syntax DAG (i.e., the left and/or right child of inner nodes): if variable $l_{i, j}$ ($r_{i, j}$) is set to $\mathit{true}$, then $j$ is the identifier of the left (right) child of node $i$.
By convention, we ignore the variables $r_{i, j}$ if node $i$ of the syntax DAG is labeled with an unary operator; similarly, we ignore both $l_{i, j}$ and $r_{i, j}$ if node $i$ is labeled with an atomic proposition.
Note that in the case of $l_{i, j}$ and $r_{i, j}$, the identifier $i$ ranges from $2$ to $n$ because node~$1$ is always labeled with an atomic proposition and, hence, cannot have children. Moreover, $j$ ranges from $1$ to $i-1$ to reflect the fact that identifier of children have to be smaller than the identifier of the current node.

To enforce that the variables $x_{i, \lambda}$, $l_{i, j}$, and $r_{i, j}$ in fact encode a syntax DAG, we impose the constraints listed in Table~\ref{tab:sat-encoding:dag}.
Formula~\eqref{formula:syntax:1} ensures that each node is labeled with exactly one label.
Similarly, Formulas~\eqref{formula:syntax:2} and \eqref{formula:syntax:3} enforce that each node (except for node~$1$) has exactly one left and exactly one right child (although we ignore certain children if the node represents an unary operator or an atomic predicate).
Finally, Formula~\eqref{formula:syntax:4} makes sure that node~$1$ is labeled with an atomic proposition.

\begin{table}
	\caption{Constraints enforcing that the variables $x_{i, \lambda}$ encode a syntax DAG} \label{tab:sat-encoding:dag}
	\begin{align}
		\Biggl[ \bigwedge_{1 \leq i \leq n} \bigvee_{\lambda \in \mathcal P \cup \mathcal C} x_{i, \lambda} \Biggr] \land \Biggl[ \bigwedge_{1 \leq i \leq n} \bigwedge_{\lambda \neq \lambda' \in \mathcal P \cup \mathcal C} \lnot x_{i, \lambda} \lor \lnot x_{i, \lambda'} \Biggr] \label{formula:syntax:1} \\
		\Biggl[ \bigwedge_{2 \leq i \leq n} \bigvee_{1 \leq j < i} l_{i, j} \Biggr] \land \Biggl[ \bigwedge_{2 \leq i \leq n} \bigwedge_{1 \leq j < j' < i} \lnot l_{i, j} \lor \lnot l_{i, j'} \Biggr] \label{formula:syntax:2} \\
		\Biggl[ \bigwedge_{2 \leq i \leq n} \bigvee_{1 \leq j < i} r_{i, j} \Biggr] \land \Biggl[ \bigwedge_{2 \leq  i \leq n} \bigwedge_{1 \leq j < j' < i} \lnot r_{i, j} \lor \lnot r_{i, j'} \Biggr] \label{formula:syntax:3} \\
		\bigvee_{p \in \mathcal P} x_{1, p} \label{formula:syntax:4}
	\end{align}
\end{table}

Let $\Phi_n^\mathit{DAG}$ now be the conjunction of Formulas~\eqref{formula:syntax:1} to \eqref{formula:syntax:4}. Then, one can construct a syntax DAG from a model $v$ of $\Phi_n^\mathit{DAG}$ in a straightforward manner: simply label node $i$ with the unique label $\lambda$ such that $v(x_{i, \lambda}) = \True$, designate node $n$ as the root, and arrange the nodes of the DAG as uniquely described by $v(l_{i, j})$ and $v(r_{i, j})$. 
Moreover, we can easily derive an LTL formula from this syntax DAG, which we denote by $\psi_v$.
Note, however, that $\psi_v$ is not yet related to the sample $\mathcal S$ and, thus, might or might not be consistent with it.

To enforce that $\psi_v$ is indeed consistent with $\mathcal S$, we now constrain the variables $x_{i, \lambda}$, $l_{i, j}$, and $r_{i, j}$ further. More precisely, we add for each ultimately periodic word $uv^\omega$ in $\mathcal S$ a propositional formula $\Phi_n^{u,v}$ that tracks the valuation of the LTL formula encoded by $\Phi_n^\mathit{DAG}$ (and all its subformulas) on $uv^\omega$.
The observation that enables us to do this is the following.

\begin{observation} \label{obs:ultimately-periodic-words}
Let $uv^\omega \in (2^\mathcal P)^\omega$, $\psi$ be an LTL formula over $\mathcal P$, and $k \in \mathbb N$. Then, $uv^\omega[|u| + k, \infty) = uv^\omega[|u| + m, \infty)$ with $m \equiv k \mod {|v|}$.
In addition, $V(\varphi, uv^\omega[|u| + k, \infty)) = V(\varphi, uv^\omega[|u| + m, \infty))$ holds for every LTL formula $\varphi$.
\end{observation}

Intuitively, Observation~\ref{obs:ultimately-periodic-words} states that the suffixes of a word $uv^\omega$ eventually repeat periodically.
As a consequence, the valuation of an LTL formula on a word $uv^\omega$ can be determined based only on the finite prefix $uv$ (recall that the semantics of temporal operators only depend on the suffixes of a word).
To illustrate this claim, consider the LTL formula $\X \varphi$ and assume that we want to determine the valuation $V(\X\varphi, uv^\omega[|uv|-1, \infty))$ (i.e., $\X \varphi$ is evaluated at the end of the prefix $uv$).
Then, Observation~\ref{obs:ultimately-periodic-words} permits us to compute this valuation based on $V(\varphi, uv^\omega[|u|, \infty))$, as opposed to the original semantics of the $\X$-operator, which recurs to $V(\varphi, uv^\omega[|uv|, \infty))$ (i.e., the valuation at the next position).
Note that similar, though more involved ideas can be applied to all other temporal operators. 

Each formula $\Phi_n^{u,v}$ is built over an auxiliary set of propositional variables $y_{i, t}^{u, v}$ where $i \in \{ 1, \ldots, n \}$ is a node in the syntax DAG and $t \in \{ 0, \ldots, |uv| - 1 \}$ is a position in the finite word $uv$.
The meaning of these variables is that the value of $y_{i, t}^{u, v}$ corresponds to the valuation $V(\varphi_i, uv^\omega[t, \infty))$ of the LTL subformula $\varphi_i$ that is rooted at node $i$.
Note that the set of variables for two distinct words from the sample must be disjoint.

To obtain this desired meaning of the variables $y_{i, t}^{u, v}$, we impose the constraints listed in Table~\ref{tab:sat-encoding:valuation}, which are inspired by bounded model checking~\cite{DBLP:journals/ac/BiereCCSZ03}.
Formula~\eqref{formula:semantics:1} implements the LTL semantics of atomic propositions and ensures that if node $i$ is labeled with $p \in \mathcal P$, then $y_{i, t}^{u, v}$ is set to $\True$ if and only if $p \in uv(t)$.
Next, Formulas~\eqref{formula:semantics:2} and \eqref{formula:semantics:3} implement the semantics of negation and disjunction, respectively: if node $i$ is labeled with $\lnot$ and node $j$ is its left child, then $y_{i, t}^{u, v}$ is the negation of $y_{j, t}^{u, v}$; on the other hand, if node $i$ is labeled with $\lor$, node $j$ is its left child, and node $j'$ is its right child, then $y_{i, t}^{u, v}$ is the disjunction of $y_{j, t}^{u, v}$ and $y_{j', t}^{u, v}$.
Moreover, Formula~\eqref{formula:semantics:4} implements the semantics of the $\X$-operator, following the idea of ``returning to the beginning of the periodic part $v$'' as sketched above.
Finally, Formula~\eqref{formula:semantics:5} implements the semantics of the $\U$-operator.
More precisely, the first conjunction in the consequent covers the positions $t \in \{ 0, \ldots, |u|-1 \}$ in the initial part $u$, while the second conjunct covers the positions $t \in \{ |u|, \ldots, |uv|-1 \}$ in the periodic part $v$. Thereby, the second conjunct relies on an auxiliary set $t \mathrel{\looparrowright}_{u,v} t'$ defined by
\[ t \mathrel{\looparrowright}_{u,v} t' \coloneqq \begin{cases} \{t, \ldots, t'-1\} & \text{if $t < t'$;} \\ \{ |u|, \ldots, t'-1, t, \ldots, |uv|-1 \} & \text{if $t \geq t'$,} \end{cases} \]
which contains all positions in $v$ ``between $t$ and $t'$''.
To avoid cluttering this section too much, we have omitted the description of the missing operators $\land$, $\rightarrow$, $\F$, $\G$ and the constants $\mathit{true}$ and $\mathit{false}$, which are implemented analogously.
Moreover, our SAT encoding is extensible, and additional LTL operators such as weak until or weak and strong release can easily be added.
\begin{table}
	\caption{Constraints enforcing that the variables $y_{i, t}^{u, v}$ track the valuation of the prospective LTL formula on ultimately periodic words} \label{tab:sat-encoding:valuation}
	\begin{align}
		\bigwedge_{1 \leq i \leq n} \bigwedge_{p \in \mathcal P} x_{i, p} \rightarrow \Biggl[ \bigwedge_{0 \leq t < |uv|} \begin{cases} y_{i, t}^{u, v} & \text{if $p \in uv(t)$} \\ \lnot y_{i, t}^{u, v} & \text{if $p \notin uv(t)$} \end{cases} \Biggr] \label{formula:semantics:1} \\
		\bigwedge_{\substack{1 < i \leq n \\ 1 \leq j < i}} (x_{i, \lnot} \land l_{i, j}) \rightarrow \bigwedge_{0 \leq t < |uv|} \Biggl[ y_{i, t}^{u, v} \leftrightarrow \lnot y_{j, t}^{u, v} \Biggr] \label{formula:semantics:2} \\
		\bigwedge_{\substack{1 < i \leq n \\ 1 \leq j, j' < i}} (x_{i, \lor} \land l_{i, j} \land r_{i, j'}) \rightarrow \bigwedge_{0 \leq t < |uv|} \Biggl[ y_{i, t}^{u, v} \leftrightarrow (y_{j, t}^{u, v} \lor y_{j', t}^{u, v}) \Biggr] \label{formula:semantics:3} \\
		\begin{aligned}
			\bigwedge_{\substack{1 < i \leq n \\ 1 \leq j < i}} (x_{i, \X} \land l_{i, j}) \rightarrow \\
			\Biggl[ \Biggl[ \bigwedge_{0 \leq t < |uv|-1} y_{i, t}^{u, v} \leftrightarrow y_{j, t+1}^{u, v} \Biggr] \land \Biggl[ y_{i, |uv|-1}^{u, v} \leftrightarrow y_{j, |u|}^{u, v} \Biggr] \Biggr] \label{formula:semantics:4}
		\end{aligned} \\
		\begin{aligned}
			\bigwedge_{\substack{1 < i \leq n \\ 1 \leq j, j' < i}} (x_{i, \U} \land l_{i, j} \land r_{i, j'}) \rightarrow \\
			\Biggl[ \Biggl[ \bigwedge_{0 \leq t < |u|} y_{i, t}^{u, v} \leftrightarrow \bigvee_{t \leq t' < |uv|} \Biggl[ y_{j', t'}^{u, v} \land \bigwedge_{t \leq t'' < t'} y_{j, t''}^{u, v} \Biggr] \Biggr] \land {} \\
			\Biggl[ \bigwedge_{|u| \leq t < |uv|}  y_{i, t}^{u, v} \leftrightarrow \bigvee_{|u| \leq t' < |uv|} \Biggl[ y_{j', t'}^{u, v} \land \bigwedge_{t'' \in t \mathrel{\looparrowright}_{u,v} t'} y_{j, t''}^{u, v} \Biggr] \Biggr] \Biggr]
		\end{aligned}
		\label{formula:semantics:5}
	\end{align}
\end{table}

For each $uv^\omega \in P \cup N$, let $\Phi_n^{u, v}$ now be the conjunction of Formulas~\eqref{formula:semantics:1} to \eqref{formula:semantics:5}. Then, we define
\[ \Phi_n^\mathcal S \coloneqq \Phi_n^\mathit{DAG} \land \Biggl[ \bigwedge_{uv^\omega \in P} \Phi_n^{u, v} \land y_{n, 0}^{u, v} \Biggr] \land \Biggl[ \bigwedge_{uv^\omega \in N} \Phi_n^{u, v} \land \lnot y_{n, 0}^{u, v} \Biggr]. \]
Note that the subformula $\Phi_n^{u, v} \land y_{n, 0}^{u, v}$ makes sure that $uv^\omega \in P$ satisfies the prospective LTL formula (more concretely, $uv^\omega$ starting from position $0$ satisfies the LTL formula at the root of the syntax DAG), while $\Phi_n^{u, v} \land  \lnot y_{n, 0}^{u, v}$ ensures that $uv^\omega \in N$ does not satisfy it.

To prove the correctness of our learning algorithm, we first establish that the formula $\Phi_n^\mathcal S$ has in fact the desired properties.

\begin{lemma} \label{lem:sat-formula-correct}
Let $\mathcal S = (P, N)$ be a sample, $n \in \mathbb N \setminus \{ 0 \}$, and $\Phi_n^\mathcal S$ the propositional formula defined above. Then, the following holds:
\begin{enumerate}
	\item \label{itm:lem:sat-formula-correct:1} If an LTL formula of size $n$ that is consistent with $\mathcal S$ exists, then the propositional formula $\Phi^\mathcal S_n$ is satisfiable.
	\item \label{itm:lem:sat-formula-correct:2} If $v \models \Phi^\mathcal S_n$, then $\psi_v$ is an LTL formula of size $n$ that is consistent with $\mathcal S$.
\end{enumerate}
\end{lemma}

Termination and correctness of Algorithm~\ref{alg:sat-learner} then follow from Lemma~\ref{lem:sat-formula-correct}.

\begin{theorem} \label{tmh:sat-learner-correct}
Given a sample $\mathcal S$, Algorithm~\ref{alg:sat-learner} terminates eventually and outputs an LTL formula of minimal size that is consistent with $\mathcal S$.
\end{theorem}

\begin{IEEEproof}
Since there exists a consistent LTL formula for every non-contradictory sample, Part~\ref{itm:lem:sat-formula-correct:1} of Lemma~\ref{lem:sat-formula-correct} guarantees that Algorithm~\ref{alg:sat-learner} terminates. Moreover, Part~\ref{itm:lem:sat-formula-correct:2} ensures that the output is indeed an LTL formula that is consistent with $\mathcal S$. Since $n$ is increased by one in every iteration of the loop until $\Phi_n^\mathcal S$ becomes satisfiable, the output of Algorithm~\ref{alg:sat-learner} is a consistent LTL formula of minimal size. 
\end{IEEEproof}

It is important to emphasize that the size of $\Phi_n^\mathcal S$ and, hence, the performance of Algorithm~\ref{alg:sat-learner} depends on the size of a sample $\mathcal S = (P, N)$, as summarized next.

\begin{remark}
The formula $\Phi_n^\mathcal S$ ranges over $\mathcal O(n^2 + n|\mathcal S|)$ variables and is of size $\mathcal O(n^2 + n^3 \sum_{uv^\omega \in P \cup N} |uv|^3)$.
\end{remark}

Finally. we conclude this section with a remark on incorporating expert knowledge into the learning process.

\begin{remark}
By adding constraints to the variables $x_{i, \lambda}$, $l_{i, j}$, and $r_{i, j}$, one can easily incorporate expert knowledge (e.g., syntactic templates) into the learning process.
\end{remark}

\section{A Decision Tree Based Learning Algorithm}
\label{sec:dtMethod}
The SAT-based algorithm described in Section~\ref{sec:sat} is an elegant, out-of-the-box way to discover minimal LTL formulas describing a sample. 
Even though it scales well beyond toy examples, its running time seems too prohibitive for real-world examples (as discussed in Section~\ref{sec:evaluation}). 
That is why we now present a learning algorithm based on a combination of SAT solving and decision tree learning.

Our second algorithm proceeds in two phases, outlined in Algorithm~\ref{alg:dt-learner}. 
In the first phase, we run Algorithm~\ref{alg:sat-learner} on small subsets of $P$ and $N$. 
This is repeated until we obtain a set $\Pi$ of LTL formulas (we call them \emph{LTL primitives}) that separate all pairs of words from $P$ and $N$.
In the second phase, formulas from $\Pi$ are used as features for a standard decision tree learning algorithm~\cite{DBLP:books/mk/Quinlan93}.
 The resulting decision tree is a Boolean combination of LTL formulas $\varphi_i \in \Pi$ that is consistent with the sample.
 
\begin{algorithm}
	\KwIn{a sample $\mathcal S$}
	\BlankLine
	Run Algorithm~\ref{alg:sat-learner} on small subsets of $P$ and $N$ to construct a set $\Pi = \{ \varphi_1, \ldots, \varphi_n\}$ of LTL formulas such that for each pair $u_1v_1^\omega \in P$ and $u_2v_2^\omega \in N$ there exists a $\varphi_i \in \Pi$ with $V(\varphi_i, u_1v_1^\omega) = 1$ and $V(\varphi_i, u_2v_2^\omega) = 0$\;
Learn a decision tree $t$ with LTL primitives from $\Pi$ as features and \Return the resulting Boolean combination $\psi_t$ of LTL primitives (which is consistent with $\mathcal S$)\;
	\caption{Learning algorithm based on decision trees}\label{alg:dt-learner}
\end{algorithm}

Note that this relaxes the problem addressed in Section~\ref{sec:sat}: we can no longer guarantee finding a formula of minimal size.
However, decision trees are among the structures that are the easiest to interpret by end-users. 
That makes them suitable for our use-case, and the minimality of formulas is replaced by structural simplicity of decision trees.

\paragraph*{Learning Decision Trees}
We assume familiarity with decision tree learning and refer the reader to a standard textbook for further details~\cite{DBLP:books/daglib/0087929}.
As illustrated in Figure~\ref{fig:dtExample}, the decision trees we seek to learn are tree-shaped structures whose inner nodes are labeled with LTL formulas from $\Pi$ and whose leaves are labeled with either $\mathit{true}$ or $\mathit{false}$.
The LTL formula represented by such a tree $t$ is given by $\psi_t \coloneqq \bigvee_{\rho \in \mathfrak P} \bigwedge_{\varphi \in \rho} \varphi$ where $\mathfrak P$ is the set of all paths from the root to a leaf labeled with $\mathit{true}$ and $\varphi \in \rho$ denotes that $\varphi$ occurs on $\rho$ (negated if the path follows a dashed edge).

To learn a decision tree over LTL primitives, we perform a preprocessing step and modify the sample as follows. For each word $uv^\omega \in P \cup N$, we use the LTL primitives as features and create a Boolean vector of size $|\Pi|$ with the $i$-th entry set to $V(\varphi_i, uv^\omega)$; this vector is then labeled with $\mathit{true}$ if $uv^\omega \in P$ or with $\mathit{false}$ if $uv^\omega \in N$.
In the second step, we apply a standard learning algorithm for decision trees to this modified sample (we used Gini impurity~\cite{breiman1993classification} as split heuristic in our experiments). 
Since we are interested in a tree that classifies our sample correctly, we disable heuristics such as pruning.

\paragraph*{Obtaining LTL Primitives}
Meaningful features are essential for a successful classification using decision trees. 
In our algorithm, features are generated from the set of LTL primitives $\Pi$.
We used two different strategies, called Strategy~$\alpha$ and Strategy~$\beta$, for obtaining $\Pi$.

Strategy~$\alpha$ iteratively chooses subsets $P' \subset P$ and $N' \subset N$ of size $k$ according to probability distributions $\text{prob}_P$ and $\text{prob}_N$ on $P$ and $N$, respectively. 
After a formula $\varphi$ separating $P'$ and $N'$ is found using Algorithm~\ref{alg:sat-learner} and added to $\Pi$, $\text{prob}_P$ and $\text{prob}_N$ are updated to increase the likelihood of any word that is not yet classified correctly by any of the $\varphi \in \Pi$ to be selected. 
This process is repeated until all pairs of positive and negative examples are separated by some LTL primitive or restarted after a user-given number of iterations.
Although this strategy is, in general, not guaranteed to terminate due to its probabilistic nature, it always did in our experiments.

Strategy~$\beta$ computes LTL primitives in a more aggressive way.
Starting with the set $S = P \times N$, it uniformly at random selects $k$ pairs from $S$ and uses Algorithm~\ref{alg:sat-learner} to compute an LTL primitive $\varphi$ that separates those pairs.
Then, it removes all pairs separated by $\varphi$ from $S$ and repeats the process until $S$ becomes empty (i.e., all pairs of examples are separated).

We refer to the extended version of this paper~\cite{technicalReport} for a detailed explanation of both strategies.

\paragraph*{Correctness}
The correctness of Algorithm~\ref{alg:dt-learner} is formalized below.

\begin{theorem}\label{thm:dtSeparation}
Given a sample $\mathcal S$, Algorithm~\ref{alg:dt-learner} learns a (not necessarily minimal) formula $\psi_t$ that is consistent with $\mathcal S$.
\end{theorem}

Theorem~\ref{thm:dtSeparation} follows from the fact that Step~1 of Algorithm~\ref{alg:dt-learner} constructs a set of LTL primitives that allows separating any pair of positive and negative examples.
Once such a set is constructed, any decision tree learner produces a decision tree $t$ that is guaranteed to classify the examples correctly. The resulting LTL formula $\psi_t$, hence, is consistent with $\mathcal S$.

\section{Evaluation}
\label{sec:evaluation}
In this section, we answer questions that arise naturally: 
how performant is Algorithm~\ref{alg:sat-learner} and what is the performance gain of Algorithm~\ref{alg:dt-learner}.
Furthermore, what is the complexity of the learned decision trees in terms of the number of decision nodes, and, finally,
how do different parameters influence the performance of Algorithm~\ref{alg:dt-learner}.
After answering these questions with experiments performed on synthetic data, we demonstrate the usefulness of our algorithms for understanding executions of a leader-election algorithm.

We implemented both learning algorithms in a Python tool\footnote{Our tool is publicly available at \url{https://github.com/gergia/samples2LTL}.} using Microsoft Z3~\cite{z3MouraBjorner}.
All experiments were conducted on Debian machines with Intel Xeon E7-8857 CPUs at 3\,GHz, using up to 5\,GB of RAM.

\paragraph*{Performance on Synthetic Data}
To simulate real-world use-cases, we generated samples based on common LTL patterns~\cite{Dwyer:1998:PSP:298595.298598}, which are shown in Table~\ref{tab:ltlPatterns}.
Starting from a pattern formula $\psi$, we generated sets of random words and separated them into $P$ and $N$ depending on whether they are a model of $\psi$ or not. 
Thereby, we fixed $|u|+|v| = 10$ for all words in the sample and added noise in form of one additional atomic proposition that is not constrained by the pattern formula. 
The size of the generated samples ranges between 50 and 5000. In total, we generated 192 samples.

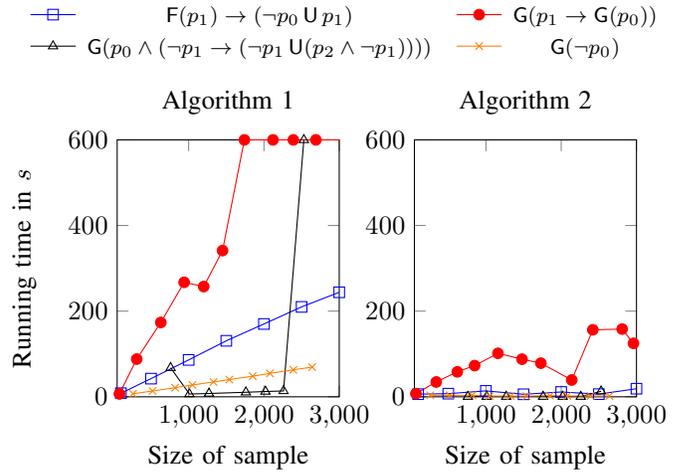
\begin{figure}
\begin{tikzpicture}
    \begin{groupplot}[group style={group size= 2 by 1},height=5cm,width=0.25\textwidth, xmin=50, xmax=3000, ymin=0, ymax=600]

        \nextgroupplot[title=Algorithm~\ref{alg:sat-learner},ylabel={Running time in $s$ }, xlabel={Size of sample}]
        \addplot[
    color=blue,
    mark=square,
    ]
    coordinates {
(100,8.335977447)
(500,42.44848162)
(1000,86.11942547)
(1500,130.7701903)
(2000,170.0726865)
(2500,209.9510419)
(3000,244.0311845)
(3500,295.0637203)
(4000,356.8050514)
(4500,395.5378082)
(5000,439.2928944)    };
    \label{plots:plot1}

               \addplot[
    color=red,
    mark=*,
    ]
    coordinates {
    (80,6.82703471509739)
(310,88.0879674439784)
(630,173.598524470115)
(940,267.343608523951)
(1200,257.423457985045)
(1450,341.522939770016)
(1740,600)
(2120,600)
(2390,600)
(2690,600)
(3010,600)

    };
\label{plots:plot2}
               \addplot[
    color=black,
    mark=triangle,
    ]
    coordinates {
(760,67.67177117)
(1010,6.49187164)
(1270,7.724078365)
(1760,10.60221089)
(2020,12.35647304)
(2260,13.6644979)
(2530,600)
    };\label{plots:plot3}
  \addplot[
    color=orange,
    mark=x,
    ]
    coordinates {
(260,7.067362176)
(520,13.52912098)
(820,21.30148176)
(1050,27.59454412)
(1330,34.68539774)
(1540,40.18596201)
(1850,48.15656359)
(2080,54.74402815)
(2390,62.97098414)
(2640,69.31923643)

    };\label{plots:plot4}

                \coordinate (top) at (rel axis cs:0,1);
        \nextgroupplot[title=Algorithm~\ref{alg:dt-learner}, xlabel={Size of sample}]
\addplot[
    color=blue,
    mark=square,
    ]
    coordinates {
(100,6.17017783224582)
(500,7.3466380983591)
(1000,13.8329336270689)
(1500,5.87141052260994)
(2000,11.2467828392982)
(2500,6.22950097173452)
(3000,18.8697897568345)
(3500,14.3169390596449)
(4000,18.5361933559179)
(4500,5.62444566562771)
(5000,6.0482433065772)};
  \label{plots:plot1}
    
\addplot[
    color=red,
    mark=*,
    ]
    coordinates {
   (70,7.36131700500845)
(340,34.6228296384215)
(620,58.298107728362)
(850,72.890760153532)
(1160,101.481917206197)
(1480,87.816773109138)
(1730,78.5400398820638)
(2140,39.2058507539331)
(2420,156.423550676554)
(2810,158.16319906339)
(2960,124.804289065301)
    };
  \label{plots:plot2}
 
\addplot[
    color=black,
    mark=triangle,
    ]
    coordinates {
(760,0.376287296414375)
(1010,0.370296027511358)
(1270,0.441913940012455)
(1760,0.380980629473924)
(2020,0.4102718308568)
(2260,0.367598194628953)
(2530,12.9229758493602)
    };
  \label{plots:plot3}
    
\addplot[
    color=orange,
    mark=x,
    ]
    coordinates {
(260,3.5807915367186)
(520,3.2556882686913)
(820,3.62299896776676)
(1050,1.86671976372599)
(1330,1.72113645821809)
(1540,1.60314920544624)
(1850,2.69230148568749)
(2080,2.00934483483433)
(2390,1.78284212574362)
(2640,1.88565448671579)
    };
  \label{plots:plot4}

                \coordinate (bot) at (rel axis cs:1,0);
    \end{groupplot}

    \path (top-|current bounding box.west)-- 
          (bot-|current bounding box.west);
\path (top|-current bounding box.north)--
      coordinate(legendpos)
      (bot|-current bounding box.north);
\matrix[
    matrix of nodes,
    anchor=south,
    inner sep=0.2em,
  ]at([yshift=1ex, xshift=-1em]legendpos)
  {
    \ref{plots:plot1}& \footnotesize $\F(p_1) \rightarrow (\neg p_0 \U p_1)$ &[5pt]
    \ref{plots:plot2}& \footnotesize $\G(p_1 \rightarrow \G(p_0))$&[5pt]\\
    \ref{plots:plot3}& \footnotesize $\G(p_0 \land (\neg p_1 \rightarrow (\neg p_1 \U (p_2 \land \neg p_1))))$&[5pt]
    \ref{plots:plot4}& \footnotesize $\G(\neg p_0)$\\};
\end{tikzpicture}
\caption{Comparison of Algorithm~\ref{alg:sat-learner} and Algorithm~\ref{alg:dt-learner}}
\label{fig:dtAndSat}
\end{figure}

Figure~\ref{fig:dtAndSat} compares the running times of Algorithm~\ref{alg:sat-learner} and Algorithm~\ref{alg:dt-learner} (using Strategy~$\alpha$ and $k=3$) on samples of varying sizes.
(So as not to clutter the presentation too much, we selected four LTL patterns that showed a typical behavior of our learning algorithms. The complete results are available in the technical report~\cite{technicalReport}.) 
Overall, Algorithm~\ref{alg:sat-learner} produces minimal formulas consistent with a sample.
It does so even for samples of considerable size, but if the sample size grows beyond 2000 (varies over samples), the SAT-based learner (Algorithm~\ref{alg:sat-learner}) frequently times out.
When Algorithm~\ref{alg:dt-learner} (using decision tree learning) is applied to these samples---as shown on the right-hand-side of Figure~\ref{fig:dtAndSat}---none of the computations timed out and the running times significantly improved. 

\begin{table}[t]
\caption{Common LTL patterns used in practice~\cite{Dwyer:1998:PSP:298595.298598}}
\label{tab:ltlPatterns}
\centering
\resizebox{\columnwidth}{!}{
$\begin{array}{ ccccc }
\toprule
\text{Absence} & \text{Existence} & \text{Universality}\\
\midrule
\G(\neg p_0) & \F(p_0) & \G(p_0)\\
\F(p_1) \rightarrow (\neg p_0 \U p_1) & \G(\neg p_0 \lor \F(p_0 \land \F(p_1))) & \F(p_1) \rightarrow (p_0 \U p_1)\\
\G(p_1 \rightarrow \G(\neg p_0)) & \G(p_0 \land (\neg p_1 \rightarrow (\neg p_1 \U (p_2 \land \neg p_1))))&\G(p_1 \rightarrow \G(p_0))\\
\bottomrule
\end{array}$
}
\end{table}

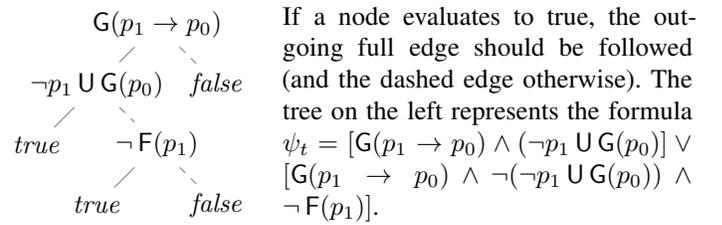
\begin{figure}[ht]
\begin{minipage}[th!]{0.20\textwidth}
	\begin{tikzpicture}
		\node (1) at (0, 0) {$\G(p_1 \rightarrow p_0)$};
		\node (2) at (-.8, -.8) {$\neg p_1 \U \G(p_0)$};
		\node (3) at (.8, -.8) {$\mathit{false}$};
		\node (4) at (-.8, -2.4) {$\mathit{true}$};
		\node (5) at (-1.6, -1.6) {$\mathit{true}$};
		\node (6) at (0, -1.6) {$\neg \F(p_1)$};
		\node (7) at (0.8, -2.4) {$\mathit{false}$};
		\draw[gray] (1) -- (2) (2) -- (5) (6) -- (4);
		\draw[gray, dashed] (1) -- (3) (2) -- (6) (6) -- (7);
	\end{tikzpicture}
\end{minipage}
\begin{minipage}[th!]{0.30\textwidth}
If a node evaluates to true, the outgoing full edge should be followed (and the dashed edge otherwise). 
The tree on the left represents the formula $\psi_t =[ \G(p_1 \rightarrow p_0) \land (\neg p_1 \U \G(p_0)] \lor [\G(p_1 \rightarrow p_0)  \land  \neg(\neg p_1 \U \G(p_0))\land \neg \F (p_1)]$.
 
 \end{minipage}
\caption{A decision tree obtained from a sample generated from the LTL pattern $\G(p_1 \rightarrow \G(p_0))$}
 \label{fig:dtExample}
 \end{figure}

What kind of trees does Algorithm~\ref{alg:dt-learner} produce? An example output of the algorithm is shown in Figure~\ref{fig:dtExample}. 
Moreover, as Table~\ref{tab:dtParams} illustrates, Algorithm~\ref{alg:dt-learner} learns small trees, often with less than five inner nodes. 
Upon closer inspection, we noticed that it often happens that one of the LTL primitives was the specified formula itself.
This suggests that small subsets already characterize our samples completely.
 
To be able to compare decision trees to the formulas learned by Algorithm~\ref{alg:dt-learner}, we measure the \emph{size} of a tree $t$ in terms of the size of the formula $\psi_t$ this tree encodes.
In our experiments, the formulas learned by Algorithm~\ref{alg:dt-learner} were on average 1.41 times larger than those learned by Algorithm~\ref{alg:sat-learner}.
However, there are outlier trees that are four times bigger than the one learned by Algorithm~\ref{alg:sat-learner}. Nonetheless, about 70\,\% are of the same size.
Even for the outliers, as emphasized previously, the readability does not degrade completely because the rule-based structure of decision trees is known to be easily understandable by humans.
Note that the runtime and size of decision trees depends on the parameters of Algorithm~\ref{alg:dt-learner}, which we discuss next.

\paragraph*{Tuning the Decision Tree-Based Algorithm}
As described in Section~\ref{sec:dtMethod}, Algorithm~\ref{alg:dt-learner} can be tuned by various parameters (sampling strategy for obtaining LTL primitives, size of sample subsets, probability increase rate, and number of repetitions inside a single sampling).
In this subsection, we explore how those parameters affect the performance of the algorithm.

 \begin{table}[h!]
\caption{Different parameters used for Algorithm~\ref{alg:dt-learner}}
\label{tab:dtParams}
\resizebox{\columnwidth}{!}{
\begin{tabular}{LLLLL}
\toprule
Sampling strategy&Subset size $k$ &Number of timeouts& Avg.\ running time in s &Avg.\ number of nodes in a tree\\
\midrule
$\alpha$ & 3 &0 / 192&21.00 &3.05\\
$\alpha$& 6&4 / 192&35.28& 1.47\\
$\alpha$& 10&8 / 192&42.72& 1.2\\
\midrule
$\beta$& 3& 4 / 192& 30.92 & 1.37 \\
$\beta$ & 6 & 12 / 192 & 48.46 &1.19 \\
$\beta$ & 10 & 21 / 192 & 48.11 &1.06 \\
\bottomrule
\end{tabular}
}
\end{table}

Table~\ref{tab:dtParams} shows the performance of Algorithm~\ref{alg:dt-learner} for different parameters, averaged over all $192$ benchmarks.
As the table indicates, the less aggressive method of separating sets, Strategy~$\alpha$, performs better. 
It seems that if the subset sizes are increased, or Strategy~$\beta$ is used, the sampled subsets already describe the specified formula completely. 
Finally, we chose Strategy~$\alpha$ and $k=3$ to be our default parameters. Varying the probability decrease rate and the number of repetitions inside a single sampling did not influence the performance much.

\paragraph*{Explaining Executions of a Leader Election Protocol}
A number of methods exist for finding errors or reproducing certain behavior in distributed systems through systematic testing~\cite{DBLP:conf/asplos/BurckhardtKMN10,DBLP:journals/pacmpl/MajumdarN18}. However, finding an execution and a corresponding schedule is only a first step towards understanding an issue. 
In the following, we demonstrate how to apply our technique in order to obtain a minimal LTL description of a specific inconsistency in a leader election protocol.

The leader election protocol we consider is the \textit{Fast Leader Election} algorithm~\cite{DBLP:conf/dsn/JunqueiraRS11,zookeeperLEExplained} used by Apache Zookeeper.
In this protocol, every node has a unique ID and initially tries to become the leader.
To this end, every node sends messages to all other nodes proclaiming its leadership. 
Upon receiving a message by an aspirant leader with a higher ID, a node gives up its claim and acknowledges its support for the aspirant.
If a node learns that an aspirant node has a support of a majority of all nodes, it commits (after waiting for a constant time for new messages) to the aspirant as the leader. 
Once committed, the node never again changes its decision and informs any other node of its commitment (one example is the message depicted by the dotted arrow in Figure~\ref{fig:LEBuggy}). 
If a node has not committed and learns about another node that has committed, it commits to the same leader. 

\begin{figure}[!th]
\centering
\begin{minipage}[t]{0.23\textwidth}
\resizebox{\columnwidth}{5cm}{
	\tikzstyle{line} = [draw]
	\tikzstyle{arrow} = [draw, -latex']
	\begin{tikzpicture}
		\node (1) at (0, 0) {\textbf{node 1}};
		\node (0) at (-2, 0) {\textbf{node 0}};
		\node (2) at (2, 0) {\textbf{node 2}};			
		\node (4) at (0, -1) {aspirant};
		\node (3) at (-2, -1) {aspirant};
		\node (5) at (2, -1) {aspirant};			
	
		\node (6) at (-2, -2) {supporting 1};
		\node (7) at (-2, -5) {supporting 2};	
		\node (8)[align=left] at (0, -3) {majority\\secured};
 		\node (9)[align=left] at (0, -4) {supporting 2};		
 		\node (14)[align=left] at (2, -5) {majority\\secured};	
 		\node (10)[align=left] at (0, -6) {\textbf{committed 2}};			
 		\node (12)[align=left] at (2, -7) {\textbf{committed 2}};		
	 	\node (13) [align=left] at (-2, -7) {\textbf{committed 2}};	

		\draw (3) -- (6);	
		\draw (6) -- (7);	
		\draw (4) -- (8);	
		\draw (8) -- (9);	
		\draw (9) -- (10);	
		\draw (5) -- (14);	
		\draw (7) -- (13);	
		\draw (14) -- (12);
		
		\draw[->] (4) -- node [above, sloped] {\emph{P1}} (6);
		\draw[->] (5)  edge [bend left=20]  node [above, sloped] {\emph{P2}} (9);
		\draw[->] (6) -- node [above, sloped] {\emph{A1}} (8);	
	    \draw[->] (9) -- node [above, sloped] {\emph{P2}} (7);	
    	\draw[->] (9) -- node [above, sloped] {\emph{A2}} (14);	
	\end{tikzpicture}
}
\caption{\small Consistent schedule for an execution of the leader election protocol}
\label{fig:LECorrect}
\end{minipage}
\hfill
\begin{minipage}[t]{0.23\textwidth}
\centering
\resizebox{\columnwidth}{5cm}{
	\begin{tikzpicture}
		\node (1) at (0, 0) {\textbf{node 1}};
		\node (0) at (-2, 0) {\textbf{node 0}};
		\node (2) at (2, 0) {\textbf{node 2}};			

		\node (4) at (0, -1) {aspirant};
		\node (3) at (-2, -1) {aspirant};
		\node (5) at (2, -1) {aspirant};			
		\node (14) at (2, -4.5) {aspirant};			
		\node (6) at (-2, -2) {supporting 1};
		\node (7) at (-2, -5) {supporting 2};	
		\node (15) at (-2, -7) {\textbf{committed 2}};	
		\node (8)[align=left] at (0, -2.5) {majority\\secured};
	 	\node (9)[align=left] at (0, -3.5) {committed 1};			
	 	\node (16)[align=left] at (0, -4.5) {\textbf{committed 1}};			
	 	\node (12)[align=left] at (2, -6) {majority\\secured};			
  		\node (13)[align=left] at (2, -7) {\textbf{committed 2}};			
	
		\draw (3) -- (6);	
		\draw (6) -- (7);	
		\draw (4) -- (8);	
		\draw (8) -- (9);	
		\draw (9) -- (16);		
		\draw (7) -- (15);
		\draw (5) -- (14);	
		\draw (14) -- (12);
		\draw (12) -- (13);	

		\draw[->] (4) -- node [above, sloped] {\emph{P1}} (6);
		\draw[->] (6) -- node [below, sloped] {\emph{A1}} (8);	
		\draw[->] (14) edge[bend left=10] node [below, sloped] {\emph{P2}} (7);	
		\draw[->] (7) edge[bend right=30] node [below, sloped] {\emph{A2}} (12);	
	  	\draw[->] (5) edge[bend left=30, dashed] node [above, sloped] {\emph{P2}} (16);	
	\end{tikzpicture}
}
\caption{\small Inconsistent schedule for an execution of the leader election protocol}
\label{fig:LEBuggy}
\end{minipage}\hfill
\end{figure}

Figure~\ref{fig:LECorrect} shows an example of a successful leader election with three nodes in an UML-style message sequence chart.
The messages exchanged between nodes are proposing the leader~$i$  ($P_i$) and node~$j$ acknowledging the claim of a leader ($A_j$). 
The arrows indicate exchanged messages and imply a precedence of events.
Note that not all messages are shown in the figures, but only the ones important for understanding the protocol.

In Figure~\ref{fig:LECorrect} all the nodes have committed to the same leader.
On the other hand, Figure~\ref{fig:LEBuggy} shows a schedule that ends up in an inconsistent state where nodes committed to different leaders. 
This schedule was discovered by the PCTCP algorithm~\cite{OzkanNiksicPCTCP}, which systematically explores the space of possible executions of distributed algorithms. 
The situation in Figure~\ref{fig:LEBuggy} is caused by the asynchronous communication:
for performance reasons, nodes commit as quickly as possible and then discard any messages, which otherwise would have changed their commitment (indicated as a dashed line in Figure~\ref{fig:LEBuggy}).
Note, however, that this is not a bug in Zookeeper's broadcast algorithm, as a leader without a quorum will not be allowed to perform any action in the later phase.

To better understand how this inconsistent state arises, our goal is to generate an LTL formula that describes the difference between the schedules in Figures~\ref{fig:LECorrect} and~\ref{fig:LEBuggy}. 
To this end, we constructed a sample by generating 20 linearizations of the schedule from Figure~\ref{fig:LECorrect} and 20 linearizations of the schedule from Figure~\ref{fig:LEBuggy}.
Since we seek an explanation for the inconsistent behavior, the former (with consistent outcomes) correspond to negative examples (set $N$), and the latter (with inconsistent outcomes) correspond to positive examples (set $P$).
 The set of atomic propositions used to construct the examples contains twelve elements: $\textit{\text{recv}}(i, j)$ for $i,j \in \{1, 2, 3 \}$ (meaning that node $j$ received a message from node $i$) and $\textit{\text{comm}}(i)$ for $i\in\{1,2,3\}$ (meaning that node $i$ committed to a leader).\footnote{While we could have included more information into propositions, we had to obscure some in order to avoid ``stating the obvious'' of the form ``node~1 committed to node~1 as a leader, while node~2 committed to node~2''.}

Finally, we ran Algorithm~\ref{alg:sat-learner} on this sample. 
The result was the formula $\neg{\textit{\text{recv}}}(2, 1) \U \textit{\text{comm}}(1)$. 
Intuitively, node~1 did not receive a message from node~2 before it committed to a leader. 
 That is exactly the difference between the schedules in Figures~\ref{fig:LECorrect} and~\ref{fig:LEBuggy}. 
 Also, it hints at a specific reason for the inconsistency in Figure~\ref{fig:LEBuggy}, thus potentially helping the engineers improve the system.
Note, however, that this experiment still required a significant amount of manual effort.
In order to apply the technique in practice, more automation is needed.

\paragraph*{Summary} 
Algorithm~\ref{alg:dt-learner} significantly improves upon the performance of Algorithm~\ref{alg:sat-learner}, though with a small increase in the size of the formula. 
The original motivation of getting readable explanations for the behavior of a system is preserved due to the fact that decision-trees are easy to comprehend.
Algorithm~\ref{alg:dt-learner} works the best using Strategy~$\alpha$ and subsets of size $k=3$.
Finally, our techniques are able to give interesting insight into real-world systems.

\section{Conclusion}
We have presented two novel algorithms for learning LTL formulas from examples. 
Our first algorithm is based on SAT solving, while the second algorithm extends the first with techniques for learning decision trees.
We have shown that both algorithms are able to learn LTL formulas for a comprehensive set of benchmarks that we have derived from common LTL patterns.
Moreover, we have demonstrated how our methods can help understand distributed algorithms.

Interesting directions of future work include the integration of LTL past-time operators, lifting our techniques to an active learning setup~\cite{DBLP:journals/iandc/Angluin87}, as well as the development of similar learning algorithms for CTL. 
Furthermore, we plan to investigate the use of maximum-margin classifiers, such as support vector machines. To this end, one needs to develop a notion of distance between temporal formulas and words, which is clearly of independent, theoretical interest as well.

\bibliographystyle{IEEEtran}
\bibliography{references}

\end{document}